\documentstyle[12pt]{article}
\newcommand{\tr}{\mbox{Tr}}

  \let\la=\lambda 
    \let\s=\sigma
 
  \let\PH=\Phi

\def\0{\over } \def\1{\vec }     \def\2{{1\over2}} \def\4{{1\over4}}
\def\5{\bar }  \def\6{\partial } \def\7#1{{#1}\llap{/}}
\def\8#1{{\textstyle{#1}}}       \def\9#1{{\bf {#1}}}
 \def\llp{\hbox to 0pt{\hss /\hskip1.5pt}}
\def\llo{\hbox to 0.2pt{\hss /}} \def\llq{\hbox to 0pt{\hss /\hskip0.5pt}}
\def\so{\supset\hbox to 0pt{\hss $\displaystyle -$\hskip1pt}}

\def\<{\langle } \def\>{\rangle }

   \let\hc=\dagger

\let\nn=\nonumber
\def\bea{\begin{eqnarray}} \def\eea{\end{eqnarray}}
\def\beann{\begin{eqnarray*}} \def\eeann{\end{eqnarray*}}
\def\beq{\begin{equation}} \def\eeq{\end{equation}}

\date{}
\title{
{\large\rm DESY 94-202}\hfill{\large\tt ISSN 0418-9833}\\
{\large\rm November 1994}\hfill\vspace*{3.0cm}\\
Phase Structure and Phase Transition\\
of the SU(2) Higgs Model in Three Dimensions}
\author{W. Buchm\"{u}ller and O. Philipsen\thanks{Address after 30
September 1994: Theoretical Physics, 1 Keble Road, Oxford OX1 3NP, UK}\\
{\normalsize\it Deutsches Elektronen-Synchrotron DESY, 22603 Hamburg, Germany}
\vspace*{3.0cm}\\
}
\begin{document}

\setlength{\baselineskip}{18pt}
\maketitle
\begin{abstract}
\thispagestyle{empty}
\noindent
We derive a set of gauge independent gap equations for Higgs boson and
vector boson masses for the SU(2) Higgs model in three dimensions.
The solutions can be associated with the Higgs phase and
the symmetric phase, respectively. In the Higgs phase the calculated masses
are in agreement with results from perturbation theory. In the symmetric
phase a non-perturbative vector boson mass is generated
by the non-abelian gauge interactions, whose value
is rather independent of the scalar self-coupling $\lambda$.
For small values of $\lambda$ the phase transition
is first-order. Its strength decreases with increasing $\lambda$,
and at a critical value $\lambda_c$ the first-order transition changes
to a crossover. Based on a perturbative matching the
three-dimensional theory is related to the four-dimensional theory
at high temperatures. The critical Higgs mass $m_H^c$,
corresponding to the critical coupling $\lambda_c$,
is estimated to be below 100 GeV. The ``symmetric phase'' of the
theory can be interpreted as a Higgs phase whose parameters are
determined non-perturbatively.
The obtained Higgs boson and vector boson masses are
compared with recent results from lattice Monte
Carlo simulations.
\end{abstract}
\setcounter{page}{0}
\newpage
\section{Introduction}
An important aspect of the standard model of strong and electroweak
interactions is the prediction of a phase transition at high temperatures,
where the electroweak symmetry is restored \cite{kirzh}. This transition
is a direct consequence of the Higgs mechanism of electroweak symmetry
breaking. It is of great cosmological importance because baryon-number
violating processes come into thermal equilibrium as the temperature
approaches the critical temperature of the transition \cite{kuzmin}.
As a consequence, the present value of the
baryon asymmetry of the universe has finally been
determined at the electroweak transition\footnote{For a recent
review, see \cite{cohen}}.

The dynamics of the electroweak phase transition has recently been studied
in detail by means of perturbation theory \cite{dine}-\cite{hebecker}
and lattice Monte Carlo simulations \cite{bunk}-\cite{karsch}.
As a first step towards the treatment of the full standard model,
the pure SU(2) Higgs model is usually investigated, neglecting
the effects of fermions and the mixing between photon and neutral vector boson.
So far we know that the transition
is weakly first-order for Higgs masses $m_H$ small compared to the
vector boson mass $m_W$. Already several years ago
it has been shown by lattice simulations that
the transition is consistent with a crossover
for very large Higgs masses \cite{evertz}.
However, for Higgs masses $m_H = {\cal O}(m_W)$
the strength, and even the nature of the transition
are not yet known. Since the mass of the physical Higgs boson may
very well be close to or larger than the W-boson mass,
it is crucial to improve our understanding
of the transition in this mass range.

At high temperatures the SU(2) Higgs model in four dimensions
can be approximated by
an effective three-dimensional theory \cite{appel}. In fact, the order
of the transition and the properties of the symmetric phase are essentially
determined by the quanta with Matsubara frequency zero, i.e., by
the three-dimensional theory. Also, the size of non-perturbative effects is
related to the confinement scale of the three-dimensional
gauge theory \cite{reuter}-\cite{wetter} and, based on confinement
in three dimensions, it has been suggested that the transition between
the Higgs phase and the symmetric phase can only be of first-order
or a crossover\cite{reuter}. Hence, one may hope to gain insight into the
nature of the phase transition at large Higgs masses by exploring directly
the Higgs model in three dimensions. The connection between this theory
and the high-temperature expansion of the Higgs model in four dimensions
has already been investigated in detail in perturbation theory
\cite{patkos},\cite{farakos}.

In this paper we shall attempt to study some non-perturbative aspects of the
Higgs model in three dimensions by means of gap equations.
In the Higgs phase as well as in the symmetric phase one expects
non-zero masses for both, the vector boson and the Higgs boson.
On the contrary, in perturbation theory
the vector boson mass vanishes at any finite order in the symmetric phase.
This suggests to perform an improved loop expansion
using masses which are self-consistently determined.
In a similar way, finite-temperature
perturbation theory requires a resummation of plasma mass effects.

Our guiding principle in the derivation of gap equations for Higgs boson
and vector boson masses will be the preservation of gauge invariance.
As we shall see, this requires a vertex resummation in addition
to the mass resummation in a unique way. The resulting
gap equations have solutions which can be associated with the Higgs phase
and the symmetric phase, respectively. Using perturbative matching
equations, which relate the three-dimensional Higgs model and the
four-dimensional Higgs model in the high temperature limit, we can
then study the implications of our results for the electroweak transition.

The paper is organized as follows. In sect.~2 some general aspects
of the SU(2) Higgs model are discussed. In sect.~3
gauge independent gap equations are derived, whose solutions are
described in sect.~4. Sect.~5 deals with implications for the
electroweak transition, and our results are summarized in sect.~6.

\section{General properties of the Higgs model}

The SU(2) Higgs model in three dimensions is defined by the action
\beq\label{l3d}
S = \int d^3x \; \tr \left[{1\over 2}W_{\mu\nu}W_{\mu\nu} +
(D_{\mu}\PH)^\hc D_{\mu}\PH + \mu^2 \PH^\hc \PH
+ 2 \lambda (\PH^\hc \PH)^2 \right] \, ,
\eeq
with
\beq
\PH = {1\over 2} (\s + i \vec{\pi}\cdot \vec{\tau}) \, ,\quad
D_{\mu}\PH = (\partial_{\mu} - i g W_{\mu})\PH\, ,\quad
W_{\mu} = {1\over 2}\vec{\tau}\cdot \vec{W_{\mu}}\ .
\eeq
Here $\vec{W_{\mu}}$ is the vector field, $\s$ is the Higgs field, $\vec{\pi}$
is the Goldstone boson field and $\vec{\tau}$ is the triplet of Pauli matrices.
The gauge coupling $g$ and the scalar coupling
$\lambda$ have mass dimension 1/2 and 1, respectively. For perturbative
calculations gauge fixing and ghost terms have to be added.

The model is known to have only one phase\footnote{For a general
discussion and references, see \cite{momu}}. However, depending on
the values of the parameters the physical properties of the model
may be rather different.
For some range of parameters it is
meaningful to distinguish a ``Higgs phase'' from a ``symmetric phase''.
Varying $\mu^2/g^4$ one then expects a phase transition which,
at least for small values of $\lambda/g^2$, should be of first order.

The physical content of the theory is contained in the properties of
correlation functions of gauge invariant operators. Consider the
composite vector and scalar fields
\beq
\tilde{W}_{\mu}^a(x) = \tr\left[\PH^\hc(x)
D_{\mu}\PH {\tau^a\over 2}(x)\right]\ ,\,
\tilde{\s}(x) =  \tr\left[\PH^\hc(x)\PH(x)\right] \ .
\eeq
The expectation value $<\tilde{\s}>$ plays the role of an ``order parameter'',
which can distinguish between the two phases. A transition from the
symmetric phase to the Higgs phase is characterized by an increase of
$<\tilde{\s}>$.

Other important parameters are the
Higgs boson and vector boson masses in both phases. They determine the
exponential falloff of the corresponding two-point functions at large
separation $|x-y|$,
\bea\label{expo}
<\tilde{\s}(x)\tilde{\s}(y)> \sim e^{-M |x-y|}\ ,\nn\\
<\tilde{W}(x)\tilde{W}(y)> \sim e^{-m |x-y|}\ .
\eea
In the Higgs phase, after fixing a gauge, these 2-point functions can
be evaluated in perturbation theory. One shifts the scalar field
$\PH$ around a gauge dependent vacuum expectation value, and the
masses $m$ and $M$ are then given by the exponential falloff of the
gauge dependent 2-point functions of the fields $\s'(x)$ and
$W_{\mu}^a(x)$. In momentum-space these masses are given by poles
of the corresponding propagators, which are gauge independent,
contrary to ``masses'' defined at zero momentum \cite{rebhan}.

One expects that in the Higgs phase, for $\mu^2/g^4 < 0$ and
$\lambda/g^2 < 1$, the masses $M$ and $m$ can be accurately calculated
using ordinary perturbation theory in $g$ and $\lambda$.
On the other hand, these masses cannot be evaluated perturbatively
in the symmetric phase, in particular near
$\mu^2 = 0$. At a first-order transition one expects a jump in
both masses. The masses in the symmetric phase may be ${\cal O}(g^2)$,
as in the pure gauge theory in three dimensions \cite{teper}. However,
they may also be smaller, especially near a
critical point, where the properties of the Higgs model can
be expected to be rather different from the pure gauge theory.

In principle, the Higgs mass and the vector boson mass can be
directly measured in both phases by means of lattice Monte Carlo simulations.
In practice, this becomes difficult if a large ratio of masses occurs
which could be the case near a critical point. In any case, it appears
desirable to gain some insight into the structure of the symmetric phase
also by means of analytical methods. In the past, gap equations have often
been a useful tool to estimate a non-perturbative mass gap. In the
following we shall apply this approach to the Higgs model in three dimensions.
This theory may be easier to solve than the four-dimensional theory
since the couplings have positive mass dimension. Furthermore, it is
conceivable that the Higgs model has a simpler structure
than the pure gauge theory since Higgs phase and
symmetric phase are analytically connected.

\section{The gap equations}

Our starting point is perturbation theory in the Higgs phase. Hence,
we shift the Higgs field $\s$ around
its vacuum expectation value $v$, $\s = v + \s'$, and supplement the
lagrangian (\ref{l3d}) by gauge fixing and ghost terms,
\beq\label{fix}
L_{GF} = {1\over 2\xi} (G^a)^2\ ,\qquad L_{FP} = - c^{*a}M^{ab}c^b\ .
\eeq
Here $M^{ab}$ is the variation of $G^a$ under a gauge transformation,
$\delta G^a = M^{ab} \Lambda^b$, and $c^{*a},c^a$ are the ghost fields.
We will perform our calculations in
$R_{\xi}$-gauge, which corresponds to the choice
\beq
G^a = \partial_{\mu}W^a_{\mu} + \xi {g\over 2}v\pi^a\ .
\eeq
The complete lagrangian then reads explicitly,
\bea\label{lexplit}
L &=& {1\over 4}\vec{W}_{\mu\nu}\vec{W}_{\mu\nu} + {1\over 2\xi}
(\partial_{\mu}\vec{W}_{\mu})^2 + {g^2\over 8}v^2\vec{W}_{\mu}^2 \nn\\
&& + {1\over 2}(\partial_{\mu}\s')^2 + \lambda v^2 \s'^2
   +{1\over 2}(\partial_{\mu}\vec{\pi})^2 + \xi{g^2\over 8}v^2\vec{\pi}^2\nn\\
&& +{g^2\over 4}v\s'\vec{W}_{\mu}^2 + {g\over 2}\vec{W}_{\mu}\cdot(
    \vec{\pi}\partial_{\mu}\s'-\s'\partial_{\mu}\vec{\pi})
   + {g\over 2}(\vec{W}_{\mu}\times\vec{\pi})\cdot\partial_{\mu}\vec{\pi}\nn\\
&& + {g^2\over 8}\vec{W}_{\mu}^2(\s'^2+\vec{\pi}^2)
   + \lambda v\s'(\s'^2+\vec{\pi}^2)+{\lambda\over 4}(\s'^2+\vec{\pi}^2)^2\nn\\
&& + \partial_{\mu}\vec{c^*}\partial_{\mu}\vec{c}
   + \xi {g^2\over 4}v^2 \vec{c^*}\vec{c}\nn\\
&& + g\partial_{\mu}\vec{c^*}\cdot(\vec{W}_{\mu}\times\vec{c})
   + \xi{g^2\over 4}v\s'\vec{c^*}\vec{c}
   + \xi{g^2\over 4}v\vec{c^*}\cdot(\vec{\pi}\times\vec{c})
   + {1\over 2}\mu^2 v^2 + {1\over 4}\lambda v^4\nn\\
&& + {1\over 2}(\mu^2+\lambda v^2)(\s'^2+\vec{\pi}^2)
   + v(\mu^2 +\lambda v^2)\s'\ .
\eea
The last two terms arise from the scalar part of the lagrangian (\ref{l3d})
after the shift in the Higgs field $\s$. For $\mu^2 < 0$, they vanish if
one expands around the classical minimum $v^2 = -\mu^2/\lambda$. In general,
however, these terms have to be kept.

{}From eq. (\ref{lexplit}) one reads off the propagators for vector boson,
Goldstone boson, ghost and Higgs boson, respectively,
\bea\label{prop}
D^{ab}_{\mu\nu}(p) &=& \delta_{ab}\left[
D_T(p)\left(\delta_{\mu\nu}-{p_{\mu}p_{\nu}\over p^2}
\right) + D_L(p){p_{\mu}p_{\nu}\over p^2}\right]\ ,\nn\\
D_T(p) &=& {1\over p^2+m_0^2}\ ,\,  D_L(p) = {\xi\over p^2+\xi m_0^2}\ ,\nn\\
\Delta^{ab}_{\pi}(p) &=& \Delta^{ab}_c (p) =
{\delta_{ab}\over p^2+\xi m_0^2}\ ,\nn\\
\Delta_{\s}(p) &=& {1\over p^2 + M_0^2} \ ,
\eea
with the tree level masses
\beq
m_0^2 = {g^2\over 4}v^2\ ,\ M_0^2 = \mu^2 + 3\lambda v^2.
\eeq

In perturbation theory the vacuum expectation value vanishes in the
symmetric phase, $v = 0$. This
implies that the vector boson mass vanishes at tree level, $m_0 = 0$.
It is generally expected that in the symmetric phase a non-zero vector boson
mass ${\cal O}(g^2)$ is generated non-perturbatively.
Since the loop expansion in three dimensions generates
a series in powers of $g^2/m_0$, ordinary perturbation theory with a vanishing
vector boson mass $m_0$ appears to be seriously deficient in the
symmetric phase.

A non-vanishing vector boson mass can be obtained from a coupled set of gap
equations for Higgs boson and vector boson masses as follows.
The tree level masses $m_0^2$ and $M_0^2$ are expressed as
\beq\label{masses}
m_0^2 = m^2 - \delta m^2\ ,\ M_0^2 = M^2 - \delta M^2\ ,
\eeq
where $m$ and $M$ enter the propagators of the loop expansion, and
$\delta m^2$ and $\delta M^2$ are treated perturbatively as counter terms.
In $R_{\xi}$-gauge the tree-level ghost and Goldstone boson masses
are given by $\sqrt{\xi} m_0$, where $m_0$ is the tree-level vector
boson mass. Correspondingly, we define $\sqrt{\xi} m$ as resummed ghost
and Goldstone boson mass.
One then obtains the coupled set of gap equations for Higgs boson
and vector boson masses,
\bea\label{gaps}
\delta m^2 + \Pi_T(p^2 = -m^2, m, M, \xi) = 0\ ,\nn\\
\delta M^2 + \Sigma(p^2 = -M^2, m, M, \xi) = 0\ ,
\eea
where $\Pi_T(p^2)$ is the transverse part of the vacuum polarization tensor,
\beq
\Pi_{\mu\nu}^{ab}(p) = \delta_{ab}\left[\left(\delta_{\mu\nu}-
{p_{\mu}p_{\nu}\over p^2}\right)\Pi_T(p^2) +
{p_{\mu}p_{\nu}\over p^2}\Pi_L(p^2)\right]\ .
\eeq
In the gap equations the self-energy corrections are evaluated on the mass
shell. This yields the physical screening lengths (cf. eq. (\ref{expo})),
and therefore a gauge independent result \cite{rebhan}.
The one-loop self-energy contributions $\Pi_{\mu\nu}^{ab}(p)$
and $\Sigma(p)$ are
given by the graphs shown in figs.~(1) and (2), respectively.

To obtain gauge independent masses from gap equations is a non-trivial
task.
In fact, the ``magnetic mass'' \cite{bfhw},\cite{espinosa},
which has been derived from
gap equations in the high-temperature expansion, is gauge dependent.
Our calculation shows that the mass resummation has to be supplemented
by a vertex resummation. This is not unexpected
since various 2-point, 3-point and 4-point couplings of the
lagrangian (\ref{lexplit}) are related by gauge invariance.

Let us consider the necessary vertex resummations in detail. First,
the vertices appearing in the graphs fig.~(1a)-(1d) only involve the
gauge coupling $g$. The gauge dependent terms of these graphs cancel
among themselves. In order to obtain a gauge independent result
for the contributions (1e)-(1l)
all cubic vertices linear in $\s'$ and $\pi^a$ are rewritten as
\beq \label{v3w}
{g^2 v\over 2} = g m - \delta V_{\phi\phi\phi}^g \ ,\ \phi = W,\ c,\
\pi^a,\ \s\ ,
\eeq
and the terms cubic in $\s'$ and $\pi^a$ are resummed as
\beq \label{v3s}
\lambda v = {g M^2\over 4 m} - \delta V_{\phi\phi\phi}^{\lambda}\ , \
\phi = \s',\ \pi^a\ .
\eeq
The explicit calculation shows that these resummations are
necessary and sufficient
to obtain a gauge independent result for $\Pi_T(p)$.
A gauge independent result for $\Sigma(p)$ is only obtained if also the
scalar self-coupling is resummed,
\beq \label{v4}
\lambda = {g^2 M^2\over 8 m^2} - \delta V_{\phi\phi\phi\phi}^\lambda\ ,\,
\phi = \s'\ ,\ \pi^a\ .
\eeq
Combining equations (\ref{masses}) and (\ref{v3w})-(\ref{v4}),
the lagrangian (\ref{lexplit}) takes the form,
\bea\label{lresum}
L &=& L_R + L_1 + L_0\ ,\nn\\
L_R &=& {1\over 4}\vec{W}_{\mu\nu}\vec{W}_{\mu\nu} + {1\over 2\xi}
(\partial_{\mu}\vec{W}_{\mu})^2 + {1\over 2}m^2\vec{W}_{\mu}^2 \nn\\
&& + {1\over 2}(\partial_{\mu}\s')^2 + {1\over 2}M^2\s'^2
   +{1\over 2}(\partial_{\mu}\vec{\pi})^2 + {\xi\over 2}m^2\vec{\pi}^2\nn\\
&& +{g\over 2}m\s'\vec{W}_{\mu}^2 + {g\over 2}\vec{W}_{\mu}\cdot(
    \vec{\pi}\partial_{\mu}\s'-\s'\partial_{\mu}\vec{\pi})
   + {g\over 2}(\vec{W}_{\mu}\times\vec{\pi})\cdot\partial_{\mu}\vec{\pi}\nn\\
&& + {g^2\over 8}\vec{W}_{\mu}^2(\s'^2+\vec{\pi}^2)
   + {g\over 4}{M^2\over m}\s'(\s'^2+\vec{\pi}^2)
   + {g^2\over 32}{M^2\over m^2}(\s'^2+\vec{\pi}^2)^2\nn\\
&& + \partial_{\mu}\vec{c^*}\partial_{\mu}\vec{c}
   + \xi m^2 \vec{c^*}\vec{c}\nn\\
&& + g\partial_{\mu}\vec{c^*}\cdot(\vec{W}_{\mu}\times\vec{c})
   + \xi{g\over 2}m\s'\vec{c^*}\vec{c}
   + \xi{g\over 2}m\vec{c^*}\cdot(\vec{\pi}\times\vec{c})\ ,\nn\\
L_1 &=& - \delta m^2 \left({1\over 2}\vec{W}_{\mu}^2 + {\xi\over 2}\vec{\pi}^2
     + \xi \vec{c^*}\vec{c}\right) - {1\over 2}\delta M^2 \s'^2
     + {1\over 2}(\mu^2+\lambda v^2)\vec{\pi}^2 \nn\\
&& + v(\mu^2 +\lambda v^2)\s'
   - \delta L_{\phi\phi\phi}-\delta L_{\phi\phi\phi\phi} \ ,\nn\\
L_0 &=& {1\over 2}\mu^2 v^2 + {1\over 4}\lambda v^4\ .
\eea
Here $\delta L_{\phi\phi\phi}$ and $\delta L_{\phi\phi\phi\phi}$ denote
the difference between tree level and resummed cubic and quartic vertices.
In the resummed perturbation theory only vertices from $L_R$ contribute
at one-loop order. In higher orders also the vertices from $L_1$ have to
be taken into account, like counter terms in ordinary perturbation theory.

Starting from the lagrangian $L_R$,
it is straightforward to evaluate the one-loop self-energy contributions
for vector boson and Higgs boson. The corresponding graphs are shown in
figs.~(1) and (2). Vertices with full bubbles denote resummed vertices,
and lines with full bubbles
represent resummed propagators, which are obtained from eq. (\ref{prop})
by replacing the tree level masses by the full masses. For the transverse
part of the vacuum polarization tensor we then obtain the result
\bea \label{vacpol}
\Pi_T(p^2) &=& g^2 \left[{m\over g M^2} v (\mu^2 + \lambda v^2) +
\left({p^4\over 4 m^4} - {p^2\over m^2}
-{15\over 8} + {3m^2\over M^2} - {m^2\over 8 p^2} + {M^2 \over 8 p^2}
\right.\right.\nn\\
&& \quad\left.\left.
-{1\over 4 m^4 p^2}(p^2 + m^2)^2 (p^2 +(\xi - 1) m^2)\right) A_0(m^2)
+ \left({5\over 8} - {M^2\over 8 p^2}
+ {m^2 \over 8 p^2}\right) A_0(M^2) \right. \nn\\
&& \quad\left. + \left({3\over 4 m^4}(m^4 - p^4) + {1\over 4 m^4 p^2}
(p^2 + m^2)^2 (3 p^2 + (\xi - 1) m^2 )\right) A_0(\xi m^2)\right.\nn\\
&& \quad\left. - \left({p^6\over 8 m^4} - {p^4\over m^2} - 5 p^2 + 4 m^2
\right) B_0(p^2, m^2, m^2)\right.\nn\\
&& \quad\left. + \left({m^2\over 2} - {1\over 8p^2}(p^2+M^2-m^2)^2\right)
B_0(p^2,m^2,M^2)\right.\nn\\
&& \quad\left. + {1\over 8 m^4} (m^4-p^4) \left(4\xi m^2 + p^2\right)
B_0(p^2,\xi m^2, \xi m^2) \right. \nn \\
&& \quad\left. + {1\over 4 m^4p^2}(p^2+m^2)^2
\left((p^2 + (\xi-1)m^2)^2-4m^2p^2\right)B_0(p^2,m^2,\xi m^2)
\right] \ .
\eea
Similarly, we find for the Higgs boson self-energy
\bea \label{higpol}
\Sigma(p^2) &=& g^2\left[{3\over 2 g m} v (\mu^2 + \lambda v^2) +
{3\over 4 m^2}(4 m^2 -  p^2)A_0(m^2)
+ {3 M^2\over 4 m^2}A_0(M^2) \right. \nn\\
&& \quad \left. +{3\over 4 m^2}(M^2 + p^2)A_0(\xi m^2)
+ {3\over 8 m^2}(8 m^4 + 4 m^2 p^2 + p^4)
B_0(p^2,m^2,m^2) \right. \nn \\
&& \quad \left. + {9 M^4 \over 8 m^2} B_0(p^2,M^2,M^2)
+{3\over 8 m^2}(M^4-p^4)B_0(p^2,\xi m^2,\xi m^2)
   \right]\ .
\eea
Here $A_0$ and $B_0$ are the three-dimensional integrals
\bea
A_0(m^2) &=& \int {d^3 k\over (2\pi)^3}{1\over k^2 + m^2} \nn\\
B_0(p^2,m_1^2,m_2^2) &=& \int {d^3 k\over (2\pi)^3}
 {1\over (k^2 + m_1^2)((k+p)^2 + m_2^2)}\ .
\eea
The integral $A_0$ is linearly divergent. The divergence
can be cancelled by
a counter term generated by an additive renormalization of the mass
parameter $\mu^2$ in the lagrangian (\ref{lexplit}). We will remove the
divergent part of the integral by dimensional regularization.

{}From eqs. (\ref{vacpol}) and (\ref{higpol}) one reads off that the
resummed one-loop self-energy contributions are gauge independent on the
mass shell. As described above this has been achieved by supplementing
the mass resummations by vertex resummations. Why did this work?
One easily verifies that the lagrangian $L_R$ can essentially
be obtained from the gauge invariant lagrangian
\beq \label{lbrs}
L = \tr\left[{1\over 2} W_{\mu\nu}W_{\mu\nu}
+ (D_{\mu}\PH)^\hc D_{\mu}\PH - {1\over 2}M^2 \PH^{\hc}\PH
+ {g^2\over 4}{M^2\over m^2} (\PH^\hc\PH)^2 \right] \ ,\nn\\
\eeq
by shifting the Higgs field $\s = \tr[\PH]$ around the ``classical''
minimum,
\beq
\s = \s' + {2 m\over g}\ ,
\eeq
and by adding the corresponding gauge fixing and ghost lagrangians
defined by (cf. (\ref{fix}))
\beq
G^a = \partial_{\mu}W^a_{\mu} + \xi m \pi^a \ .
\eeq
The resulting lagrangian differs from $L_R$ in eq. (\ref{lresum})
only by a constant.
Hence, the lagrangian $L_R$ is invariant under BRS transformations
and we expect a gauge independent result for the position of the pole
of a propagator.

The functions $A_0$ and $B_0$ are easily evaluated, and from
eqs. (\ref{gaps}), (\ref{vacpol}) and (\ref{higpol}) one obtains the
gap equations,
\bea \label{gigaps}
m^2 &=& m_0^2 - {g z\over M} v (\mu^2 + \lambda v^2)
 + m g^2 \bar{f}(z) \ ,  \nn\\
M^2 &=& M_0^2 - {3g\over 2m} v (\mu^2 + \lambda v^2)
 + M g^2 \bar{F}(z) \ ,
\eea
where
\bea
\bar{f}(z) &=& {1\over \pi}\left[{63\over 64} \ln3 - {1\over 8} +
{1\over 32 z^3} - {1\over 32 z^2} + {1\over 8z} \right. \nn\\
&& \left.\quad + {3\over 4} z^2 - \left({1\over 64 z^4} - {1\over 16 z^2}
+ {1\over 8}\right)\ln(1+2z)\right]\ ,\label{barfw}\\
\bar{F}(z) &=& {1\over \pi}\left[{3\over 64}(4 - 3 \ln3){1\over z^2}
+ {3\over 16z} + {3\over 4}z \right.\nn\\
&& \left.\quad - \left({3\over 8}z^2 - {3\over 16} + {3\over 64z^2}\right)
\ln{2z+1\over 2z-1}\right]\ ,\label{barfs}
\eea
with $z=m/M$.
Note, that the equation for M becomes complex for $M > 2 m$, i.e., $z < 1/2$,
since in this case the Higgs boson can decay into two vector bosons. We
will therefore restrict our discussion to the mass range $M < 2 m$.
In order to find solutions of the gap equations we have to specify the
value of the vacuum expectation value $v$. This will be discussed in the
following section.

In our derivation of the gauge independent gap equations the scalar
degrees of freedom have played a crucial role. In fact, it does not
seem possible to derive a gauge independent gap equation for the pure
SU(2) gauge theory. The physical reason for this appears rather obvious.
In the Higgs model the Goldstone bosons can screen the colour of the
gauge bosons, which is not the case in the pure gauge theory. Hence,
the ground state and the spectrum of excitations can be very different
for the two theories.

\section{Phase structure in three dimensions}

In the previous section we have performed a calculation in the
``Higgs phase'', i.e., we have shifted the Higgs
field around an unspecified vacuum expectation value $v$. The value
of $v$ can be self-consistently determined from the requirement that
the vacuum expectation value of the shifted field is zero,
\beq \label{vaceq}
<\s'>\  =\ 0\ .
\eeq
This equation simply means that the sum of all tadpole contributions to the
self-energies, which are shown in figs.~(1i) - (1m) and (2i) - (2m), vanishes.
{}From eq. (\ref{vaceq}) one obtains in resummed perturbation theory at
one-loop order,
\bea\label{vacval}
v(\mu^2 + \lambda v^2) &=& - {3 \over 4}g m \left( 4 A_0(m^2) +
{M^2\over m^2} A_0(\xi m^2) + {M^2\over m^2} A_0(M^2)\right)\nn\\
&=& {3\over 16 \pi}g\left(4 m^2 + \sqrt{\xi} M^2  + {M^3\over m}\right)\ .
\eea
The vacuum expectation value $v$ of the Higgs field is not a physical
observable and expected to be gauge dependent, as in ordinary
perturbation theory.
Numerically, the gauge dependence becomes important
for large Higgs masses, $M > m$. On the other hand,
the masses obtained from the gap equations
(\ref{gigaps}) are physical observables and must therefore be
gauge independent. The weak gauge dependence induced by the
gauge dependence of $v$ has to be
cancelled by higher order contributions.
In the following we shall work in Landau gauge, $\xi = 0$.

For any solution $v$ of eq. (\ref{vacval}),
only the irreducible parts of the self-energies
contribute in the gap equations (\ref{gigaps}). From eqs. (\ref{vacpol})
and (\ref{higpol}) one easily finds ($z=m/M$),
\bea
m^2 &=& {g^2\over 4} v^2 + m g^2 f(z)\ ,\label{mir}\\
M^2 &=& \mu^2 + 3\lambda v^2 + M g^2 F(z)\, ,\label{Mir}
\eea
where the functions $f(z)$ and $F(z)$ are given by
\bea
f(z) &=& {1\over \pi}\left[{63\over 64} \ln3 - {1\over 8} +
{1\over 32 z^3} - {1\over 32 z^2} - {1\over 16z}
- {3\sqrt\xi\over 16}\right. \nn\\
&& \left.\quad - \left({1\over 64 z^4} - {1\over 16 z^2}
+ {1\over 8}\right)\ln(1+2z)\right]\ ,\\
F(z) &=& {1\over \pi}\left[-\left({3\over 32} + {9\over 64}\ln3
\right){1\over z^2} + {3\over 16}\left(1 - {3\over 2}\sqrt\xi\right){1\over z}
\right.\nn\\
&& \left.\quad - {3\over 8} z
- \left({3\over 8}z^2 - {3\over 16} + {3\over 64z^2}\right)
\ln{2z+1\over 2z-1}\right]\ .
\eea
The solutions of the gap equations depend crucially on
the properties of these functions.
They are plotted in figs.~(3) and (4). $F(z)$ is always negative, whereas
$f(z)$ has a zero at a large value $z_0$. Their asymptotic behaviour
for large z reads
\beq
f(z) \sim -{1\over 8\pi}\ln z\ ,\quad F(z) \sim -{3\over 4\pi}z\ .
\eeq
Also important is the behaviour of $f(z)$ in the vicinity of its zero
which is given by
\beq
f(z) = -{1\over 8\pi}{z-z_0\over z_0} + {\cal O}\left(
\left({z-z_0\over z_0}\right)^2\right)\ .
\eeq

{}From equations (\ref{gigaps}) and (\ref{vacval})
we can also obtain the one-loop results of
ordinary perturbation theory for vacuum expectation value and masses.
In this case, the masses $m_0 = gv/2$ and $M_0 = \sqrt{2\lambda}v$,
with the fixed ratio $z = \sqrt{g^2/8\lambda}$,
appear in the one-loop expressions,
and the vacuum expectation value $v$ is determined from the one-loop
effective potential.
One then finds (cf. eqs. (\ref{barfw}), (\ref{barfs})),
\bea
v(\mu^2 + \lambda v^2) &=& {v^2\over 4\pi}
\left({3\over 4}g^3 + 3\sqrt{2} \lambda^{3/2}
+ {3\over 2} \sqrt{\xi} \lambda g \right) \ ,\label{pertvev}\\
m^2 &=& -{g^2\over 4\lambda}\mu^2 + {g^3\over 2} v
\bar{f}\left(\sqrt{{g^2\over 8 \lambda}}\right)\ ,\\
M^2 &=& -2\mu^2 + g^2\sqrt{2\lambda}v
\bar{F}\left(\sqrt{{g^2\over 8 \lambda}}\right)\ .
\eea
{}From these equations $v$, $m$ and $M$ are easily obtained as functions
of $g$, $\lambda$ and $\mu^2$.

The coupled system of equations (\ref{vacval}) - (\ref{Mir})
for vacuum expectation value and masses can
be solved numerically. For a given value of $\lambda/g^2$ one can
determine $v/g$, $m/g^2$ and $M/g^2$ as functions of $\mu^2/g^4$. Let
us first choose a small value, $\lambda/g^2 = 1/128$. In the
Higgs phase, this correponds to $m_H \sim m_W/4$.
The result for $v/g$ is shown in fig.~(5). For
$\mu^2/g^4 < 0$, $v/g$ is large, as expected for the Higgs phase.
For large positive values of $\mu^2/g^4$,
which correspond to the symmetric phase of the theory, $v/g$ is
small but non-zero. Furthermore, $v/g$
is rather independent of $\mu^2/g^4$. Of particular interest is the region
of small positive $\mu^2/g^4$. Here one obtains two solutions, i.e., the
theory has one metastable state. The Higgs phase is generated by
quantum corrections, as in the Coleman-Weinberg mechanism of radiative
symmetry breaking \cite{cole}. The vacuum expectation value in the Higgs
phase is not continuously connected to the one in the symmetric phase.
Hence, the phase transition is first-order.
In fig.~(5) the solutions of the gap equations are also compared with the
results from ordinary perturbation theory. In the Higgs phase the
perturbative value of $v/g$ is slightly smaller than the one
obtained from the gap equations. The main difference concerns the
symmetric phase. Here, in perturbation theory  $v = 0$, whereas
the gap equations yield $0 < v/g < 1$.

Each solution $v/g(\mu^2/g^4)$ is connected with Higgs boson and vector
boson masses $M/g^2(\mu^2/g^4)$ and $m/g^2(\mu^2/g^4)$.
In fig.~(6) the results are shown and compared with perturbation theory.
The masses in the Higgs phase are in accord
with the results of perturbation theory. In the symmetric phase,
the vector boson mass is rather small and independent of $\mu^2/g^4$.
Its value is of the same order of magnitude as the ``magnetic mass''
which has previously been obtained for the symmetric phase of the SU(2) Higgs
model at high temperatures in Landau gauge
\cite{bfhw},\cite{espinosa}. Rather intriguing
is the behaviour of vector boson and Higgs boson masses
in the symmetric phase in the metastability domain. As $\mu^2/g^4$ approaches
zero, both, $m/g^2$ and $M/g^2$ tend to zero
with a fixed ratio $M/m = 1/z_0 \ll 1$, a behaviour
very different from that obtained in perturbation theory.

In the limit $\mu^2/g^4 \rightarrow 0$, the solution of
equations (\ref{vacval})-(\ref{Mir}) takes a simple form,
\bea
v &\sim& {4\pi\over 3}{\mu^2\over g^3} \quad ,\\
{z-z_0\over z_0} &\sim& -8\pi^2{\mu^2\over g^4} \quad ,\\
M &\sim& {4\pi\over 3 z_0}{\mu^2\over g^2} \quad ,
\eea
where $z_0$ is the zero of the function $f(z)$. As in the ordinary
Higgs phase, vector boson and Higgs boson masses are proportional to
the vacuum expectation value $v$,
\beq
m \sim v \ ,\quad M \sim 2\sqrt{2\lambda_R}v\ ,
\eeq
where $\lambda_R = g^2/8z_0^2$ is the resummed scalar coupling. Hence, near
the point $\mu^2 = 0$ the smallness of the Higgs mass
reflects the smallness of the resummed scalar coupling.
Note, that the suppression of the non-perturbative contribution
to the vector boson mass
and the smallness of the effective scalar coupling for small $\mu^2/g^4$
are also characteristic features of the $\epsilon$-expansion applied
to the electroweak phase transition \cite{yaffe}.

For sufficiently small values of $\lambda/g^2$ the phase transition is
first-order. However, larger values of $\lambda$ are of particular
importance because of the present lower experimental bound for the Higgs
boson mass. Let us first consider the point $\mu^2 = 0$
where the solution, which corresponds to the symmetric phase, vanishes.
For large values of $\lambda/g^2$, the solution of eqs. (\ref{vacval})
- (\ref{Mir}) takes a form similar to the one obtained for
small $\mu^2/g^4$. In both cases the ratio $m/M$ is close to the zero
of $f(z)$. One finds,
\bea
v &\sim& {1\over 12\pi}{g^3\over \lambda}\quad ,\\
{z-z_0\over z_0} &\sim& {5\over 18}{g^2\over \lambda} \quad , \\
M &\sim& {1\over 36\pi z_0}{g^4\over \lambda}\quad .
\eea
Hence, at $\mu^2 = 0$ one has $v>0$ for arbitrarily large values of
$\lambda/g^2$. This means that for a small range of $\mu^2/g^4$ near
$\mu^2 = 0$ there exist at least two solutions of the gap equations,
one of which represents a metastable state.

What is the nature of the transition for large values of $\lambda/g^2$?
In fig.~(7) the solution $v/g(\mu^2/g^4)$ of the gap equations is shown
for $\lambda/g^2 = 1/8$. In the Higgs phase, this corresponds
to $m_H \sim m_W$. Compared to fig.~(5), where
$\lambda/g^2 = 1/128$, a dramatic change has taken place. The Higgs
phase and the symmetric phase are now continuously connected. Hence,
the first-order transition has changed to a crossover!
The phase transition changes its character at a critical
coupling $\lambda_c$. At this value of $\lambda$ the two solutions
of the gap equations merge at a critical value
$\mu_c^2$. Numerically, we find $\lambda_c/g^2\approx 0.053$ and
$\mu_c^2/g^4\approx 0.054$. The nature of the transition at this endpoint
of the first-order transition line will be studied in more detail
elsewhere.

An important quantity is the vector boson mass in the symmetric phase.
A comparison of figs.~(6) and (8) shows that its value is essentially
independent of $\lambda$. In both cases the vector boson mass agrees
within $10\%$ with the mass obtained from the nonlinear $\s$-model, where
only the graphs fig.~(1a)-(1d) contribute, which yield \cite{philipsen1}
\bea
m_{SM} &=& C g^2\ ,\nn\\
C &=& {3\over 64\pi}(21 \ln 3 - 4) = 0.28 \ .
\eea
This is in accord with the intuitive picture that for large $\mu^2/g^4$
the Higgs degree of freedom should be unimportant for the vector boson mass.
As $\mu^2/g^4$ is
increased, the ratio of the masses, $z=m/M$, crosses the value $1/2$ at some
point $\mu^2_e/g^4$, and real solutions to the gap
equations cease to exist. The
position of this point, $\mu^2_e/g^4\approx 0.8$,
is essentially independent of $\la$.

The Higgs boson and vector boson masses, which we have determined in
this section, are solutions of one-loop gap equations. What can be said
about the size of corrections in higher orders of the loop expansion?
In particular one might worry about the well known infrared problem
in the symmetric phase, where the effective gauge coupling
and also the scalar coupling can become very large \cite{reuter}-\cite{wetter}.
The expansion parameter $\rho_V$ for vector loops is
(cf. \cite{arnold}, \cite{bufo}, \cite{bhw}),
\beq
\rho_V\ =\ {1\over 6\pi}{g^2\over m}\ .
\eeq
In the Higgs phase where, for $\mu^2 \approx 0$,
$v \approx 3g^3/(16\pi\lambda)$ (cf. (\ref{pertvev})), one then has
\beq
\rho_V^{Higgs}\ \approx {16\over 9} {\lambda\over g^2}\ .
\eeq
For $m_H \sim m_W$, i.e. $\lambda/g^2 \sim 1/8$, this yields
$\rho_V^{Higgs} \approx 0.22$. On the other hand, in the symmetric phase
one has
\beq \label{loopsym}
\rho_V^{sym}\ =\ {1\over 6\pi C} \approx 0.19\ ,
\eeq
which corresponds to the expansion parameter in the Higgs phase for a Higgs
mass
slightly below the vector boson mass.

Alternatively, one may estimate the effect of higher order corrections
by means of the
running coupling in $4-\epsilon$ dimensions at $\epsilon = 1$,
\beq
g^2(\mu) = {g^2\over 1 + \beta_0 g^2/\mu}\ ,
\eeq
where $\beta_0 = - 43/(48\pi^2)$ for the SU(2) gauge theory with one
doublet of Higgs fields. At the scale of the vector boson mass
in the symmetric phase one finds $g^2(m_{SM}) \approx 1.48 g^2$, i.e.,
the running coupling is still rather far away from the infrared
singularity. The Higgs boson mass is much smaller than the vector
boson mass. However, since the scalar coupling in the symmetric phase
is also suppressed like $M^2/m^2$, it is conceivable that higher order
corrections involving the Higgs field are small.

We conclude that the solution of the one-loop gap equations discussed
in this section may indeed provide a suitable starting point for a
systematic improved loop expansion in the Higgs phase as well as in
the symmetric phase. Most difficult is the region near the phase
transition point, especially for large $\lambda/g^2$. Here, Higgs
boson and vector boson masses in the symmetric phase are small, and
the one-loop results are therefore not reliable.
This also applies to the region near $\mu^2 = 0$ in the symmetric phase.
In general, a strong gauge dependence of a one-loop result can be used as an
indication for the importance of higher order corrections.

\section{The electroweak transition}

At high temperatures the SU(2) Higgs model in four dimensions can be
described by an effective three-dimensional theory. The connection between
the two theories has been discussed in detail in perturbation theory
\cite{patkos},\cite{farakos}.
For sufficiently large values of $\lambda$, the SU(2) Higgs
model in three dimensions is expected to be the appropriate effective theory.
Perturbation theory at one-loop order yields the following relations
between the parameters (cf., e.g. \cite{bufo}),
\bea\label{match}
g^2 &=& \bar{g}^2(T)T\ ,\,
\lambda = \left(\bar{\lambda}(T) - {3\over 128\pi}\sqrt{{6\over 5}}
\bar{g}^3(T) + {\cal O}(\bar{g}^4, \bar{\lambda}^2)\right) T\ ,\nn\\
\mu^2 &=& \left({3\over 16} \bar{g}^2(T) + {1\over 2}\bar{\lambda}(T)
-{3\over 16\pi}\sqrt{{5\over 6}}\bar{g}^3(T) + {\cal O}(\bar{g}^4,
\bar{\lambda}^2) \right) (T^2 - T_b^2) \ .
\eea
Here $\bar{g}$ and $\bar{\lambda}$ are the dimensionless couplings
in four dimensions, $T$ is the temperature and $T_b$ is the
``barrier temperature''.
These matching equations can be used to interpret the results of
the previous section in terms of the high-temperature theory and
to compare them with recent results obtained by means of lattice
Monte Carlo simulations \cite{kajantie},\cite{montvay}.
Note, however, that for a precise quantitative comparison between
the three-dimensional theory and the four-dimensional theory at high
temperatures the perturbative matching equations (\ref{match}) are
not sufficient \cite{mack}.

In ref.~\cite{montvay} a first-order phase transition was found for
$m_H = 49$ GeV, $m_W = 80$ GeV and $\bar{g}^2 = 0.576$, which implies
for the scalar coupling $\bar{\lambda}=0.027$. The critical
temperature was measured to be $T_c = 93.7$ GeV, and Higgs boson
and vector boson masses were determined in the symmetric phase and the
Higgs phase. In order to compare these masses with the solutions of
the gap equations, one has to determine the critical mass parameter $\mu_c^2$
where the energy densities of the two solutions in the metastability region
are equal.
In the loop expansion, the energy density is calculated
as function of couplings,
masses and vacuum expectation value. It is not clear
that a good approximation is obtained at one-loop order if the on-shell
masses are used. This problem requires further study. However, even
without knowing $\mu^2_c$, useful bounds on the masses in the Higgs phase
and the symmetric phase can be obtained by considering the edge of
the metastability range.

\renewcommand{\arraystretch}{1.5}
\begin{table}[b]
\begin{center}
\begin{tabular}{llllll}
\hline
    &Higgs phase & & &symmetric phase &   \\
    &$m/(g^2 T_c)$ &$M/m$ & &$m/(g^2 T_c)$ &$M/m$  \\ \hline\hline
gap equations & $> 0.59$ & $> 0.25$ & & $< 0.27$ & $< 0.10$ \\ \hline
lattice simulations & 0.76 - 0.93 & 0.35 & & 1.2 - 1.7 & 0.2 - 0.3 \\ \hline
\end{tabular}
\end{center}
\caption[]{\it Comparison of vector boson and Higgs boson masses
obtained from gap equations and lattice Monte Carlo simulations.
The lattice data are from ref.~\cite{montvay}.}
\end{table}

According to eqs. (\ref{match}) the parameters of the lattice simulations
correspond to $\lambda/g^2 = 0.0406$. The corresponding vector boson
and Higgs boson masses are shown in fig.~(9) as functions of $\mu^2/g^4$.
The values at the upper end of the metastability range yield lower and
upper bounds for the masses in the Higgs phase and the symmetric phase,
respectively. The dimensionless quantities
$m/(g^2 T_c)$ and $M/m$ obtained from gap equations and lattice
simulations are given in table 1.

In the Higgs phase the two approaches yield consistent results.
With respect to the symmetric phase, however, there is a severe
discrepancy. The lattice simulations indicate a spectrum of states,
such that the smallest mass in the symmetric phase is larger
than the mass in the Higgs phase -- a result which may seem surprising.
The opposite is true for the solution of the gap equations.
Here the masses in the symmetric phase are much smaller than the ones in the
Higgs phase. Since the masses are rather small, the one-loop result may not
be reliable quantitatively. However, the qualitative feature that
the masses in the symmetric phase are smaller than the ones in the
Higgs phase is in accord with the mass predictions deep in the
symmetric phase where the
loop expansion parameter is small (cf. (\ref{loopsym})).
It is conceivable that the small masses, predicted by the
gap equations, could not yet be identified by the lattice simulations.
Simulations on larger lattices should be able to resolve this
puzzle.

As discussed in the previous section, the vector boson mass in the
symmetric phase
is given to good accuracy by the contributions
shown in fig.~(1a) - (1d). This suggests that in the symmetric phase
the dynamics of the vector bosons is rather independent of the Higgs field,
and described by the effective lagrangian
\bea\label{leff}
L_{eff} &=& {1\over 4}\vec{W}_{\mu\nu}\vec{W}_{\mu\nu} + {1\over 2\xi}
(\partial_{\mu}\vec{W}_{\mu})^2 + {1\over 2}m^2\vec{W}_{\mu}^2 \nn\\
&& + {1\over 2}(\partial_{\mu}\vec{\pi})^2 + {\xi\over 2}m^2\vec{\pi}^2
   + \partial_{\mu}\vec{c^*}\partial_{\mu}\vec{c}
   + \xi m^2 \vec{c^*}\vec{c}\nn\\
&& + {g\over 2}(\vec{W}_{\mu}\times\vec{\pi})\cdot\partial_{\mu}\vec{\pi}
    + g\partial_{\mu}\vec{c^*}\cdot(\vec{W}_{\mu}\times\vec{c})
   + \xi{g\over 2}m\vec{c^*}\cdot(\vec{\pi}\times\vec{c}) .
\eea
This lagrangian can be obtained from the gauged non-linear $\s$-model by
applying the resummation procedure described in sect.~3 and by neglecting terms
${\cal O}(\pi^3,\pi^4, \pi^3 W,\ldots)$ \cite{philipsen1}.
Since the lagrangian (\ref{leff}) yields the correct
vector boson mass in the symmetric phase, it may also be an appropriate
starting point for
evaluating the rate of baryon- and lepton-number changing processes.
Hence, our results support previous work on $B+L$-violation in the
symmetric phase \cite{philipsen}.

In the previous section we found that the first-order transition changes
to a crossover at a critical value of the scalar coupling,
$\lambda_c/g^2 \approx 0.053$. For the quoted parameters from
ref.~\cite{montvay}
this correponds to a critical Higgs mass $m_H^c \sim 55$ GeV.
This value strongly depends on the size of the vector boson mass in the
symmetric phase. If, for instance,
two-loop corrections would reduce the one-loop value,
$m_{SM} = 0.28 g^2 T$, by $50 \%$ to $\bar{m}_{SM} = 0.14 g^2 T$, the
critical Higgs mass would increase to $\bar{m}_H^c \sim 100$ GeV. Hence,
from our approach we can only conclude that the change from a first-order
transition to a crossover should take place for Higgs masses below
100 GeV. This is compatible with results from lattice calculations for
the SU(2) Higgs model in four dimensions at high temperature \cite{bunk},
as well as for the reduced three-dimensional theory \cite{kajantie},
where evidence for a first-order transition at $m_H \sim m_W$ is
claimed.

Many features of the electroweak transition, as described in this section,
are similar to effects of a ``magnetic mass'', which have been
discussed previously \cite{bfhw},\cite{espinosa}. In fact,
in ref.~\cite{bfhw} a change from a first-order to a second-order transition
was predicted at $\tilde{m}^c_H = 85/\sqrt{\gamma}$ GeV for a magnetic mass
$m_{magn} = \gamma g^2 T/(3\pi)$. For $m_{magn} = 0.28 g^2 T$ the
corresponding
critical Higgs mass is $\tilde{m}_H^c = 52$ GeV. The weakness of
the electroweak phase transition for Higgs masses
$m_H = {\cal O}(m_W)$ again raises
the question of how the transition
actually takes place and how important subcritical bubbles \cite{gleiser}
are as compared to the nucleation and growth of critical bubbles.

\section{Summary and conclusions}

In the previous sections we have derived gauge independent gap equations
for vector boson and Higgs boson masses,
and we have found different solutions of
these equations whose properties depend on the dimensionless parameters
of the theory, $\lambda/g^2$ and $\mu^2/g^4$. The presence of a Higgs
doublet of scalar fields turned out to be crucial to obtain
gauge independent equations. For the pure SU(2) gauge theory no
gauge independent equation could be found.

Two kinds of solutions of the gap equations were found, with large and
small vacuum expectation values $v/g$ of the Higgs field, which could be
associated with the Higgs phase and the symmetric phase of the theory,
respectively. In the Higgs phase the obtained masses for vector boson
and Higgs boson are in agreement with ordinary perturbation theory.
In the symmetric phase, a non-perturbative vector boson
mass is generated whose value is rather independent
of $\mu^2/g^4$ and $\lambda/g^2$. Except for a small range of $\mu^2/g^4$
close to the phase transition point, Higgs boson and vector boson masses are
large enough to yield a sufficiently small loop expansion parameter.
Hence, we expect that
the properties of the symmetric phase obtained at one-loop order
will not be qualitatively changed by higher order corrections.
The obtained masses in the symmetric phase appear to be
at variance with recent measurements of lattice simulations, a puzzle
which can be resolved by considering larger lattices.

An intriguing aspect of the picture emerging from the gap equations is
the difference between the scalar couplings in the symmetric phase and the
Higgs phase, respectively. The resummed scalar coupling $\lambda_R$ in the
symmetric phase near $\mu^2 = 0$ is very small, which is related to the
large ratio of vector boson and Higgs boson masses. A similar behaviour
of the scalar coupling is suggested by the renormalization group
equations in $4-\epsilon$ dimensions.

The nature of the transition between Higgs phase and symmetric phase
depends on the value of the scalar coupling $\lambda$.
Below a critical coupling $\lambda_c$  we find a first-order transition,
which changes to a crossover at $\lambda = \lambda_c$. The precise
value of $\lambda_c$ is strongly correlated with the size of the
vector boson mass in the symmetric phase. Based on the perturbative
matching between the three-dimensional theory and the high-temperature
expansion of the four-dimensional theory the critical
Higgs mass $m_H^c$, corresponding to the critical coupling $\lambda_c$,
is estimated to be smaller than 100 GeV.

The guiding principle in the design of a resummation procedure for
masses and vertices leading to the gap equations has been the preservation
of gauge invariance. The result has a very simple interpretation. The
lagrangian with resummed masses and couplings is nothing but the
ordinary Higgs model with modified parameters, shifted around the
corresponding ``classical'' minimum. Hence, the ``symmetric'' phase
is again a Higgs phase, just with different parameters. These
parameters are determined self-consistently at one-loop order, like
the vacuum expectation value of the Higgs field in the case of
Coleman-Weinberg type symmetry breaking by radiative corrections.
Hence, no ``symmetry restoration'' takes place at high temperatures,
the vacuum expectation value
of the Higgs field is always non-zero.
This result is in accord with the known property of the Higgs model
to have only one phase.

The solutions of the gap equations provide a clear physical picture
of the symmetric phase and the transition between Higgs phase and
symmetric phase. More work is needed to prove the stability of
the one-loop results with respect to higher order corrections and
to understand the connection between the small masses obtained from the
gap equations and the mass spectrum measured with lattice simulations.
The results of these further investigations
will be of crucial importance for achieving a full understanding of the
electroweak phase transition and its implications for the cosmological
baryon asymmetry.

We are grateful to Z. Fodor, K. Jansen, A. Hebecker, M. L\"uscher,
G. Mack, A. Rebhan and M. Reuter for valuable discussions, suggestions
and comments.

\newpage
\noindent
\\
{\bf\large Figure captions}\\
\\
{\bf Fig.1} One-loop contributions to the vector boson propagator.\\
\\
{\bf Fig.2} One-loop contributions to the Higgs boson propagator.\\
\\
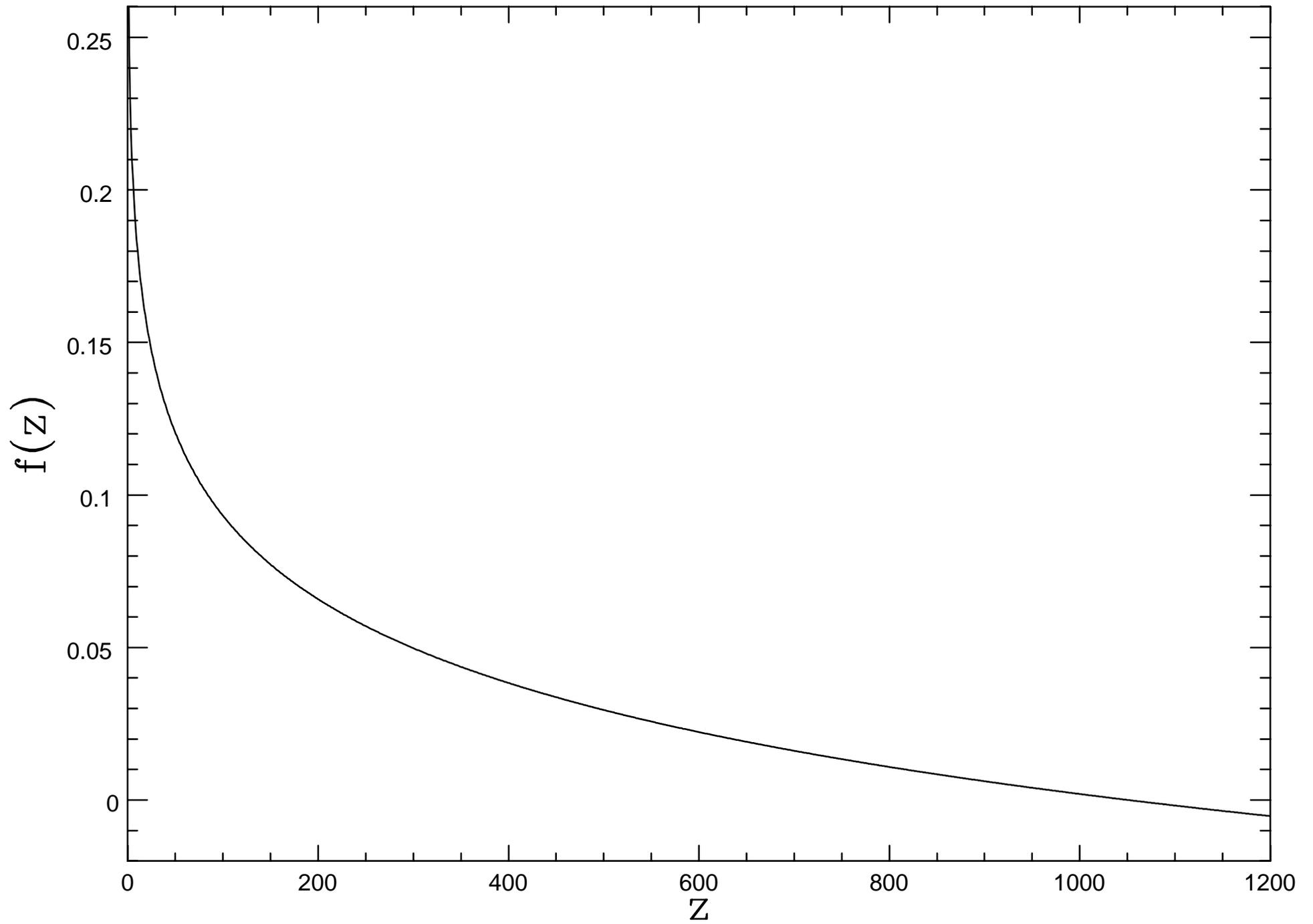
{\bf Fig.3} The function $f(z)$ entering the gap equation for the
vector boson mass, $\xi=0$.\\
\\
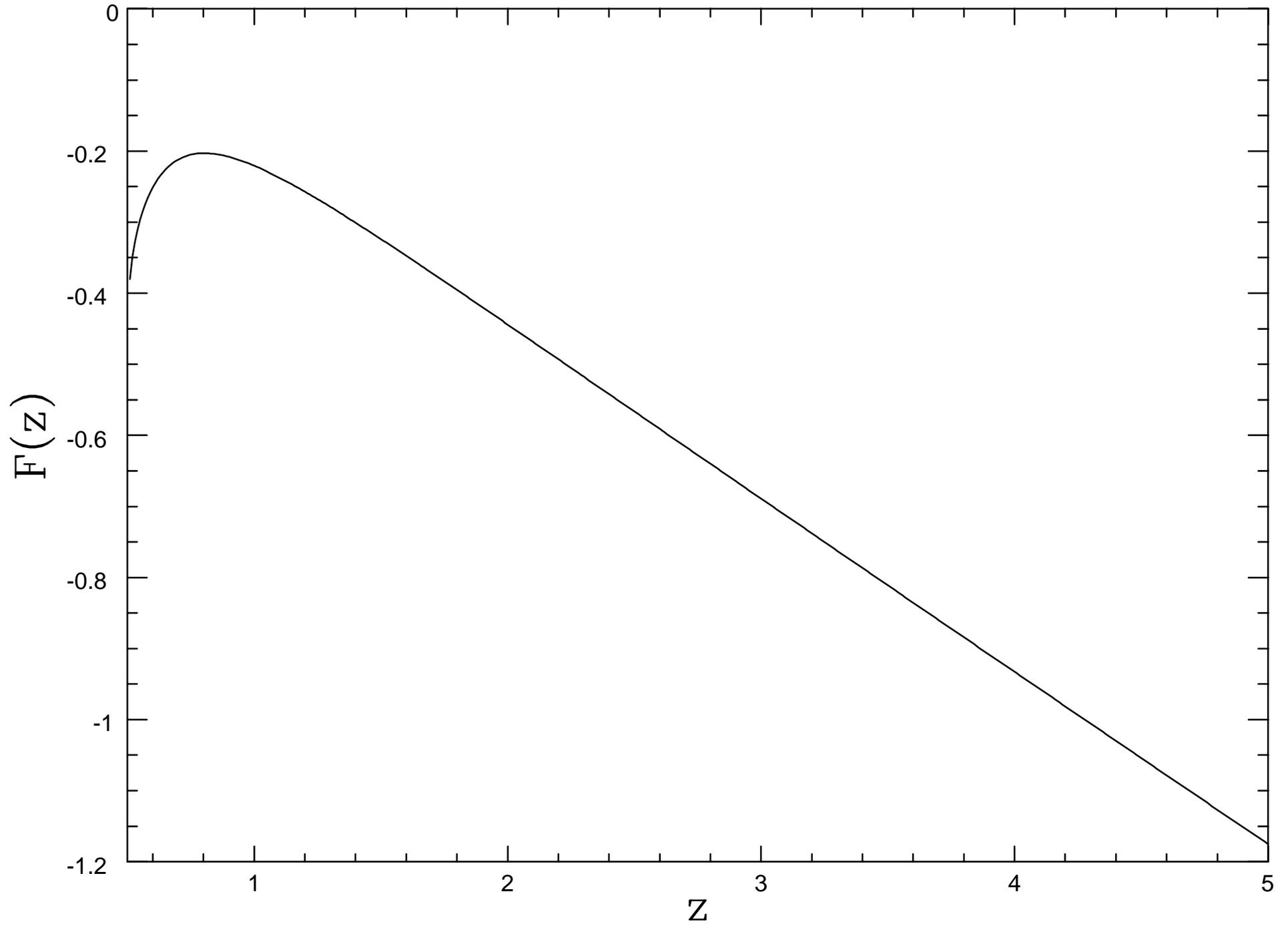
{\bf Fig.4} The function $F(z)$ entering the gap equation for the
Higgs boson mass, $\xi=0$.\\
\\
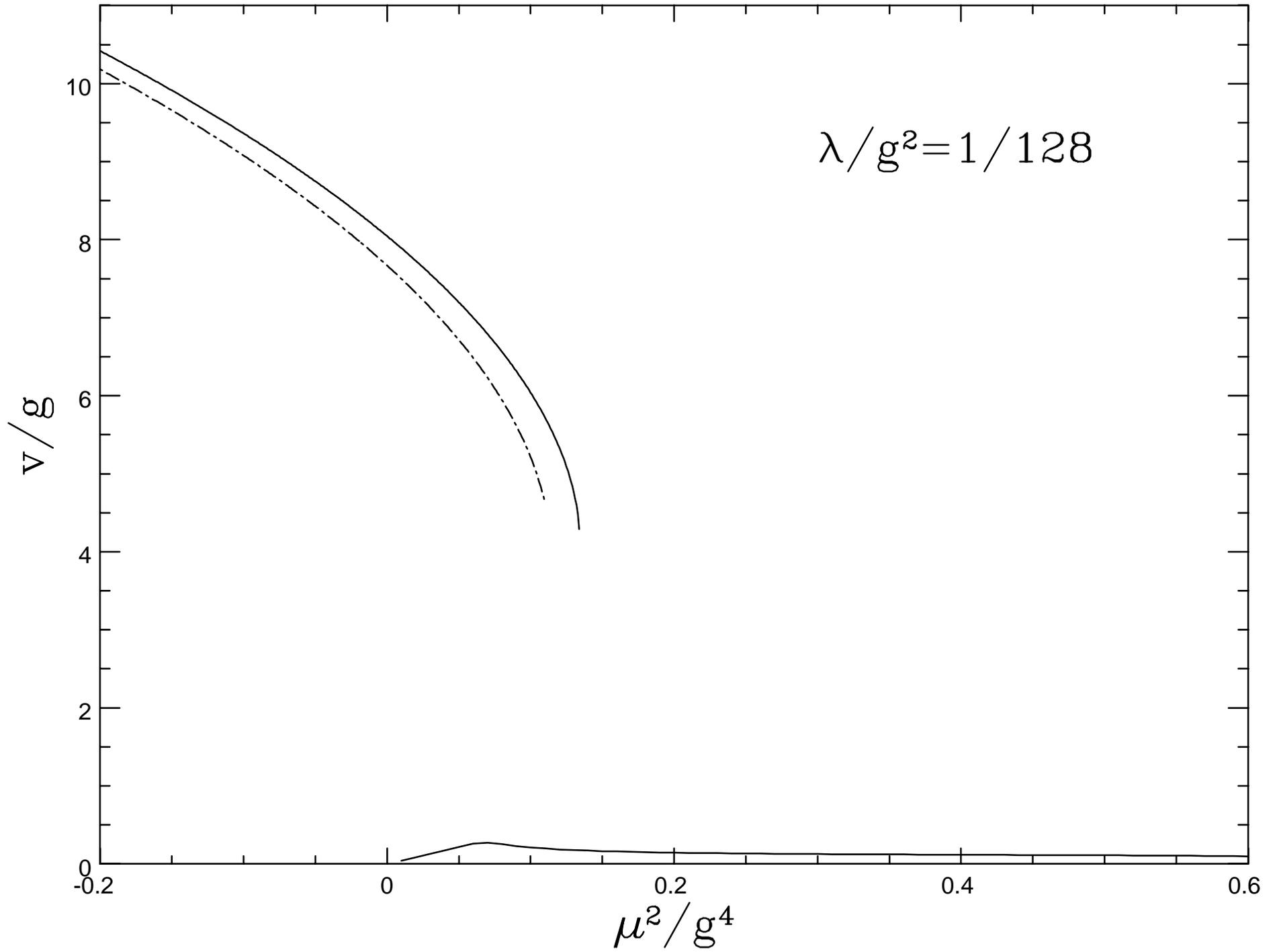
{\bf Fig.5} The vacuum expectation value $v/g$ as function of the
mass parameter $\mu^2/g^4$. Full line: solution of gap equations,
dash-dotted line: perturbation theory. $\lambda/g^2 = 1/128$.\\
\\
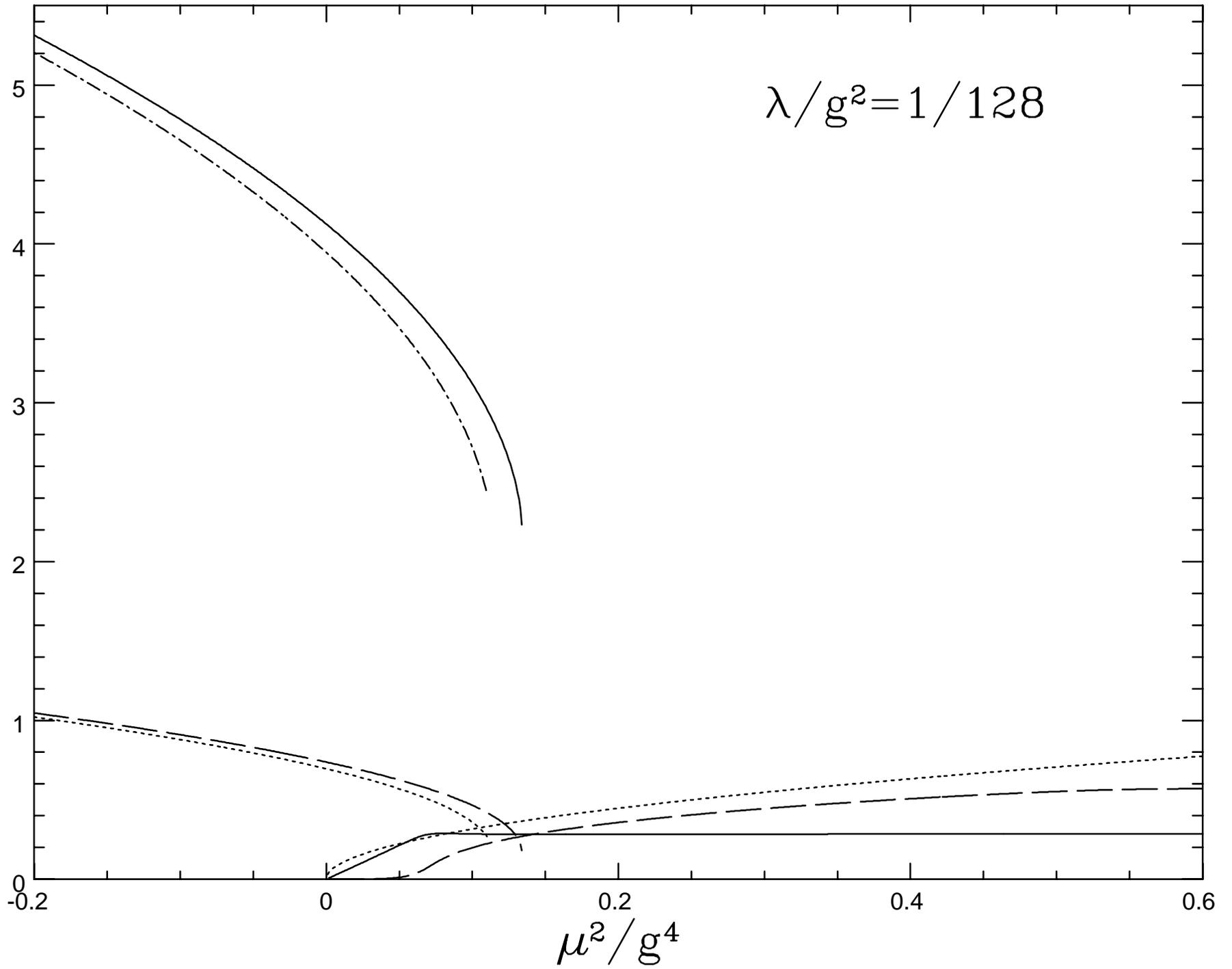
{\bf Fig.6} Vector boson and Higgs boson masses for $\lambda/g^2 = 1/128$.
Gap equations: m (full line), M (dashed line); perturbation theory:
m (dash-dotted line), M (dotted line).\\
\\
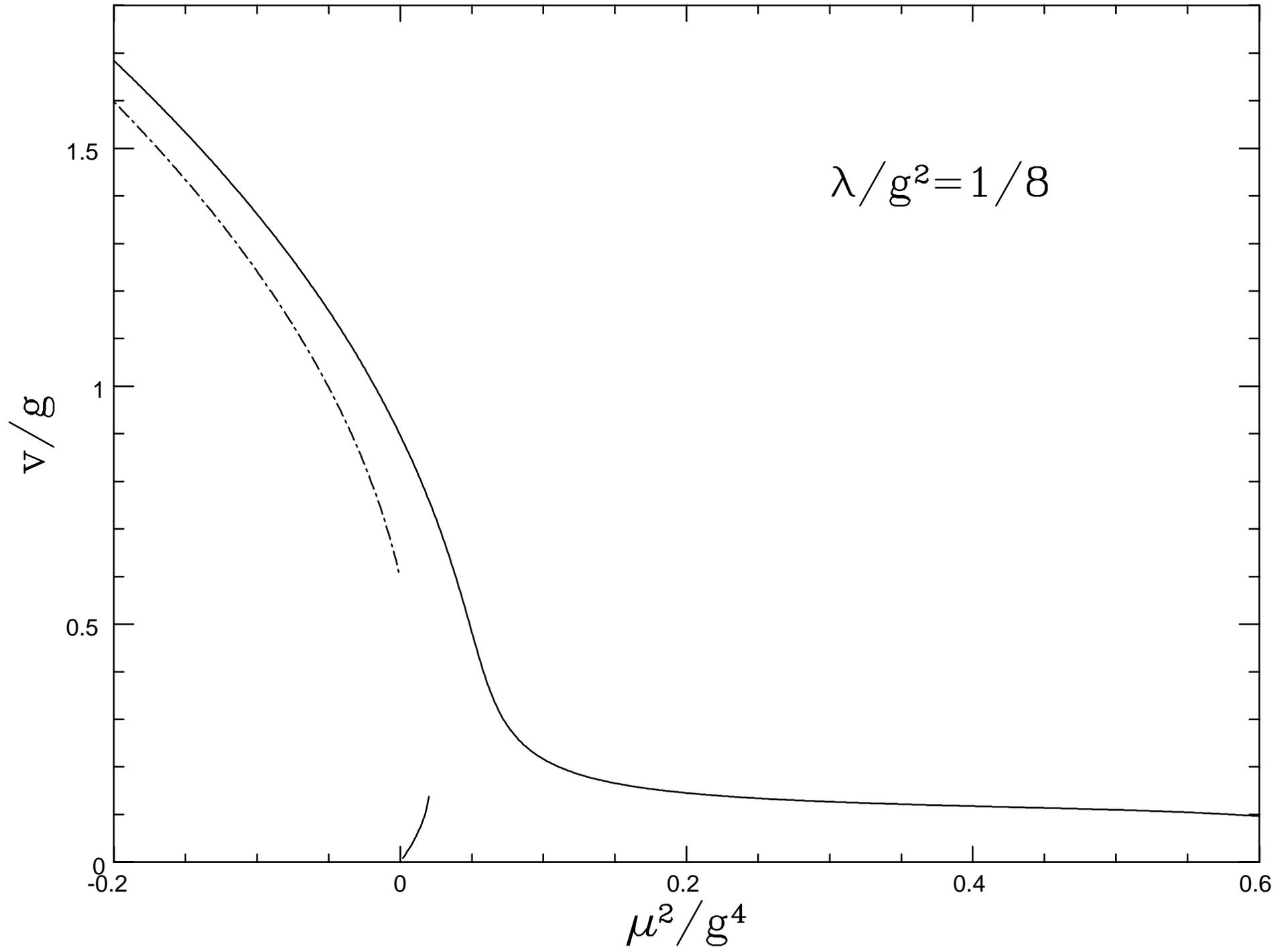
{\bf Fig.7} The vacuum expectation value $v/g$ as function of the
mass parameter $\mu^2/g^4$. Full line: solution of gap equations,
dash-dotted line: perturbation theory. $\lambda/g^2 = 1/8$.\\
\\
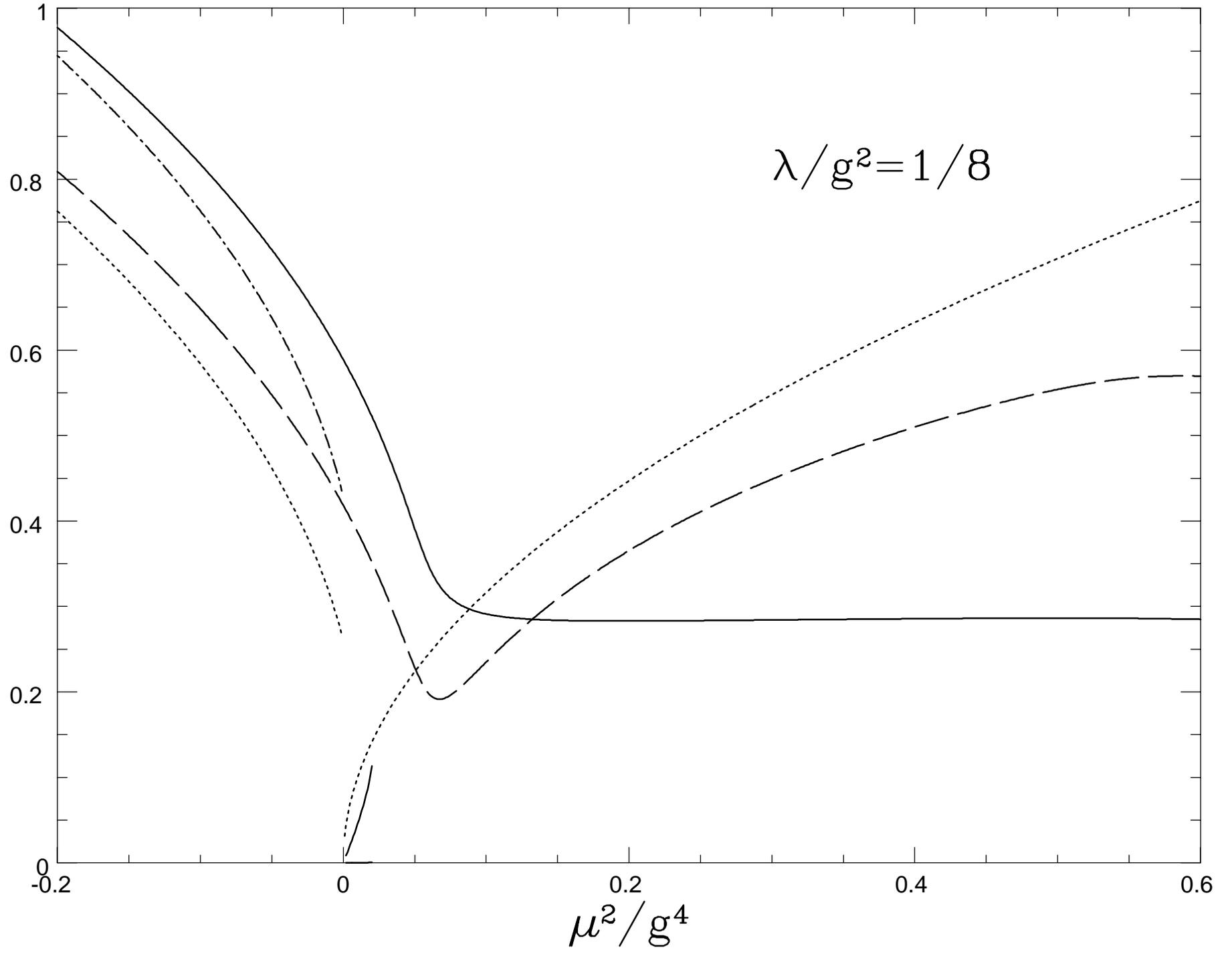
{\bf Fig.8} Vector boson and Higgs boson masses for $\lambda/g^2 = 1/8$.
Gap equations: m (full line), M (dashed line); perturbation theory:
m (dash-dotted line), M (dotted line).\\
\\
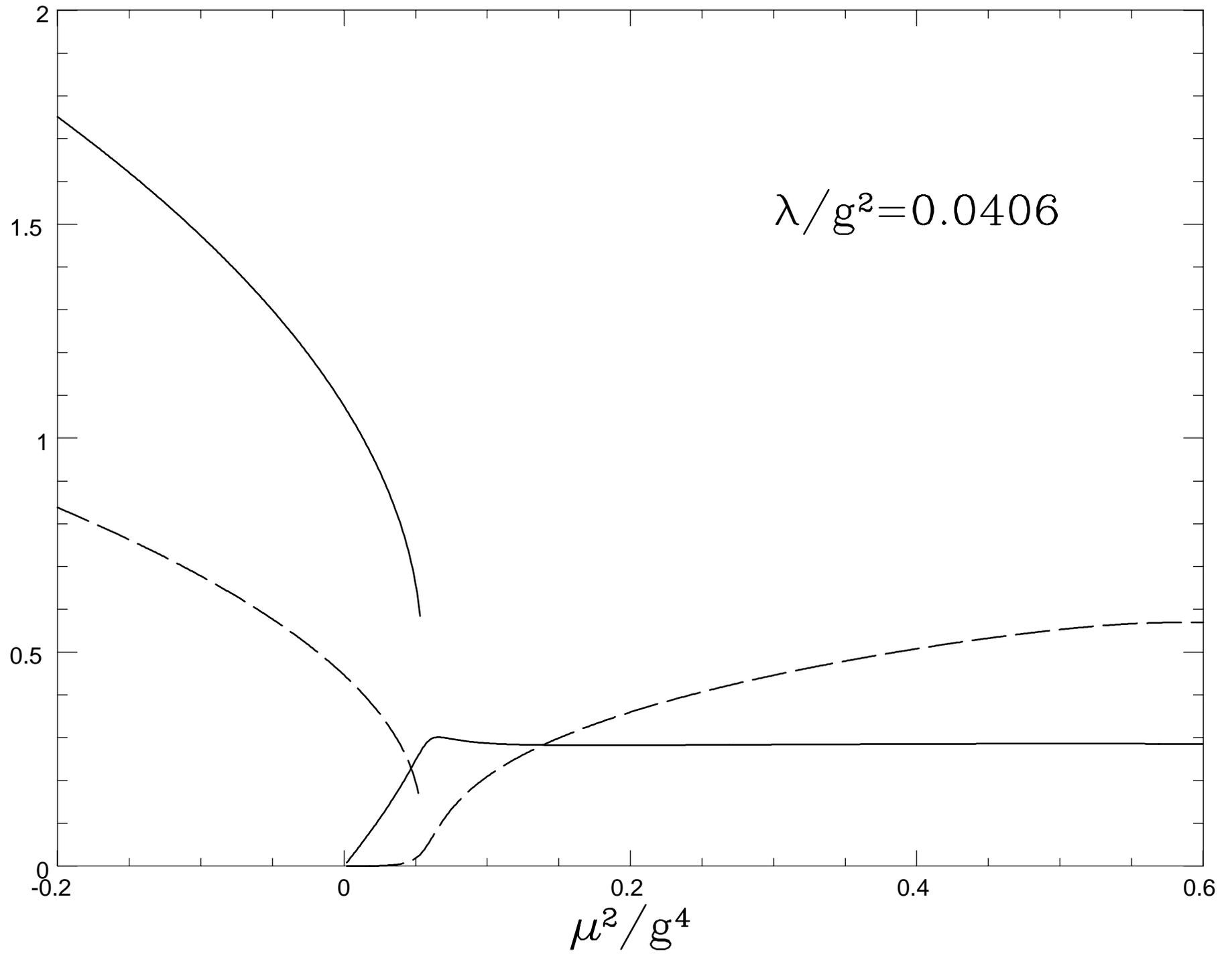
{\bf Fig.9} Vector boson mass m (full line) and Higgs boson mass M
(dashed line) for $\lambda/g^2 = 0.0406$.
\\
\end{document}